\documentclass{PoS}
\usepackage{slashed}
\title{Landau Levels in Lattice QCD}

\ShortTitle{Landau Levels in Lattice QCD}

\author{Falk Bruckmann\thanks{Supported by the DFG (BR 2872/6-1 and BR 2872/7-1)}\\
	Universit\"at Regensburg, Institut f\"ur Physik,
        Universit\"atstra\ss e 31, 93053 Regensburg, Germany\\ 
        E-mail: \email{falk.bruckmann@physik.uni-regensburg.de}}

\author{Gergely Endr\H odi\thanks{Supported by the DFG (Emmy Noether Programme EN 1064/2-1; and SFB/TRR 55)}\\
	Institute for Theoretical Physics, Goethe University,
        Max-von-Laue-Strasse 1, 
	60438 Frankfurt am Main, Germany\\
        E-mail: \email{endrodi@th.physik.uni-frankfurt.de}
}

\author{Matteo Giordano\thanks{Supported by OTKA under the grant
    OTKA-K-113034.}, S\'andor D. Katz\\ 
        Institute for Theoretical Physics, E\"otv\"os University,
and MTA-ELTE Lattice Gauge Theory Research Group,
P\'azm\'any P. s\'et\'any 1/A, H-1117 Budapest, Hungary\\
        E-mail: \email{giordano@bodri.elte.hu, katz@bodri.elte.hu}}
\author{Tam\'as G. Kov\'acs\protect{\footnotemark[3]}\mbox{ }
  \thanks{Supported by the Hungarian Academy of Sciences under
    ``Lend\"ulet'' grant No. LP2011-011.}\\
	Institute for Nuclear Research of the Hungarian Academy of Sciences,
	Bem t\'er 18/c, H-4026 Debrecen, Hungary\\
        E-mail: \email{kgt@atomki.mta.hu}
}
\author{\speaker{Ferenc Pittler}\\
	HISKP(Theory), University of Bonn,
	Nussallee 14-16, D-53115,Bonn, Germany
	E-mail:\email{pittler@hiskp.uni-bonn.de}
}
\author{Jacob Wellnhofer\protect{\footnotemark[1]}\mbox{ }\\
        Universit\"at Regensburg, Institut f\"ur Physik,
        Universit\"atstra\ss e 31, 93053 Regensburg, Germany\\ 
        E-mail:\email{Jacob.Wellnhofer@physik.uni-regensburg.de}
}
\abstract{
The spectrum of the two-dimensional continuum Dirac operator 
in the presence of a uniform background magnetic field consists 
of Landau levels, which are degenerate and separated by gaps. On 
the lattice the Landau levels are spread out by discretization 
artefacts, but a remnant of their structure is clearly visible 
(Hofstadter butterfly). If one switches on a non-Abelian interaction, 
the butterfly structure will be smeared out, but the lowest Landau 
level (LLL) will still be separated by a gap from the rest of the spectrum. 
In this talk we discuss how one can define the LLL in QCD and check how well 
certain physical quantities are approximated by taking into account
only the LLL. 
}

\FullConference{34th annual International Symposium on Lattice Field Theory\\
		24-30 July 2016\\
		University of Southampton, UK}

\begin{document}

\section{Introduction}

Strong magnetic fields play a crucial role in noncentral heavy-ion
collisions, compact stars and in  
the evolution of the early Universe. For a recent review
see~\cite{Andersen:2014xxa}. 
Progress in the study of these problems requires a better
understanding of the effect of an external magnetic field in Quantum
Chromo Dynamics (QCD), the theory which 
describes the strong interaction between  
quarks and gluons.
In QCD the external magnetic field has an impact on
the dynamical chiral symmetry breaking  
in the vacuum of the theory. Below the QCD crossover temperature the
order parameter for chiral symmetry breaking, i.e., the chiral condensate,
is enhanced by the magnetic field, a phenomenon called magnetic
catalysis~\cite{Shovkovy:2012zn}. Around the pseudo-critical
temperature the chiral 
condensate instead decreases when the magnetic field is turned on
(inverse magnetic catalysis)~\cite{Bali:2011qj}.

As the magnetic field enters the Dirac operator $\slashed{D}$, it
affects the chiral condensate in two ways~\cite{D'Elia:2011zu}. 
Since $\slashed{D}$ is the
operator of interest, there is a direct, ``valence'' effect of the
magnetic field on the observable. However, the magnetic field also
influences the probability distribution of the gauge configurations
through the fermionic determinant in the action, thus having a second, 
indirect effect on the chiral condensate. The latter we call the
``sea'' effect. In order to get some insight into the phenomena of
magnetic catalysis and inverse magnetic catalysis, it is useful to
study the valence and the sea effects separately.

The valence effect is studied by determining the spectrum of the Dirac
operator at nonzero magnetic field $B$ on a typical configuration obtained
at $B=0$. The result is an increase in the density of
the low modes of the Dirac operator, which leads through the famous
Banks-Casher relation~\cite{Banks:1979yr} to magnetic catalysis.  
However, by turning on the magnetic field also in the sea, the
fermionic determinant would suppress those gauge configurations which
contain a larger density of small eigenvalues. Thus the sea  
effect drives the system towards inverse magnetic catalysis. It turns
out that the valence effect dominates  
over the sea except around the pseudo-critical transition
temperature~\cite{Bali:2011qj}. Here the Polyakov loop   
effective potential is flat and a small effect of the magnetic field
in the fermion determinant can  
significantly change expectation values~\cite{Bruckmann:2013oba}. 



Magnetic catalysis is often attributed to the linear dependence of the
degeneracy of the lowest Landau level (LLL) on the strength of the
magnetic field. A widely employed approximation in low-energy models,
effective theories and functional approaches is to neglect higher
Landau levels, see, e.g., Refs.~\cite{Fukushima:2011nu,Leung:2005xz,
Blaizot:2012sd,  Ferrer:2014qka,Mueller:2015fka,Fayazbakhsh:2010bh,
Fukushima:2015wck,Braun:2014fua}. To assess the systematics of
such approximations, here we study the issue of Landau levels on the
lattice for the first time, and check directly their influence on the
chiral condensate.





\section{Landau levels}

In the case of free\footnote{In this context ``free'' means that the
quarks do not interact with gluons.}  
quarks exposed to a uniform background magnetic field
(which will always point in the $z$ direction), the
spectrum of the Euclidean Dirac operator is organized in so-called
Landau levels. We first examine them in $2d$, and then proceed to the
physical $(3+1)d$ case. We start with the continuum, finite volume
case, and then go to the lattice. In a finite  
periodic box of area $L^{2}$, the flux of the magnetic field 
is quantized according to
\begin{equation}
\label{quant_cond}
N_{b}\equiv \frac{qBL^{2}}{2\pi} \in \mathbb{Z}\,,
\end{equation}
and the eigenvalues of the free, massless $-\slashed{D}^2$ are
\begin{equation}
\lambda^{2}_k = \vert q B\vert k\,,
\qquad k =
2n+1-2s_{z}\mathrm{sgn}\left(qB\right)\,, 
~~ n=0,1,\ldots\,,~~
s_{z}=\pm\frac{1}{2}. 
\end{equation}
The integer $k$ identifies the Landau level (LL).  
The degeneracy $\nu_k$ of each 
LL is proportional to the magnetic flux through  
the area of the system, i.e,
\begin{equation}
\label{degen}
\nu_{k}=N_bN_c\left(2-\delta_{k,0}\right).
\end{equation}
The separation between levels is
proportional to the strength of the magnetic field. Thus,
for large magnetic fields there will be  
a huge separation between the lowest Landau level (LLL), which
(for positive $qB$) corresponds to $(n,s_{z})=(0,\frac{1}{2})$, and the rest of the
spectrum, since   
the LLL is $B$ independent. If this separation of modes persists even
after turning on the strong interaction, then one could easily explain
magnetic catalysis in 2$d$: the enhanced density 
of the lowest modes would in fact increase the condensate through the
Banks-Casher relation. 

\begin{figure}[t]
 \centering
 \includegraphics[width=7cm]{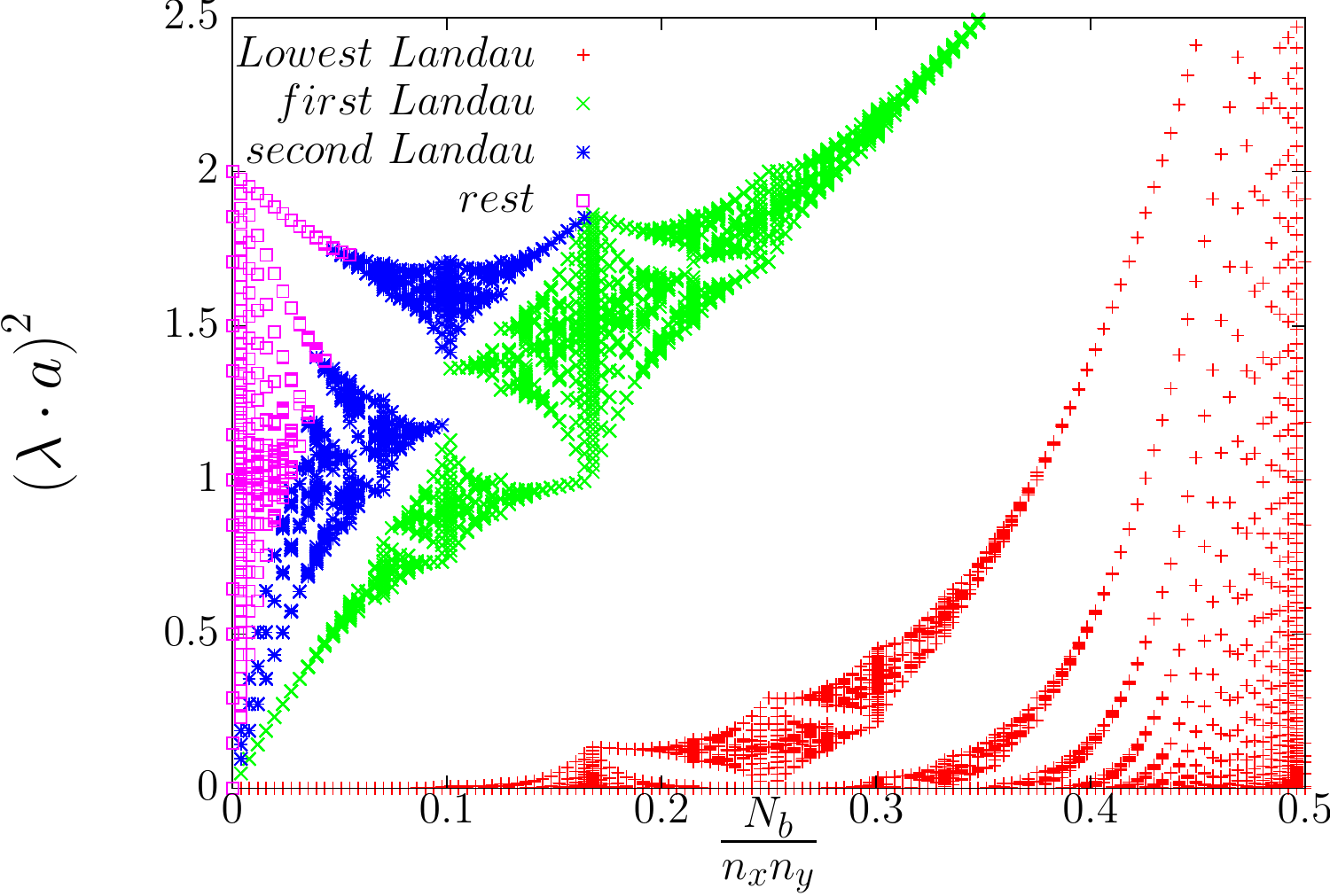}\quad\quad
 \includegraphics[width=7cm]{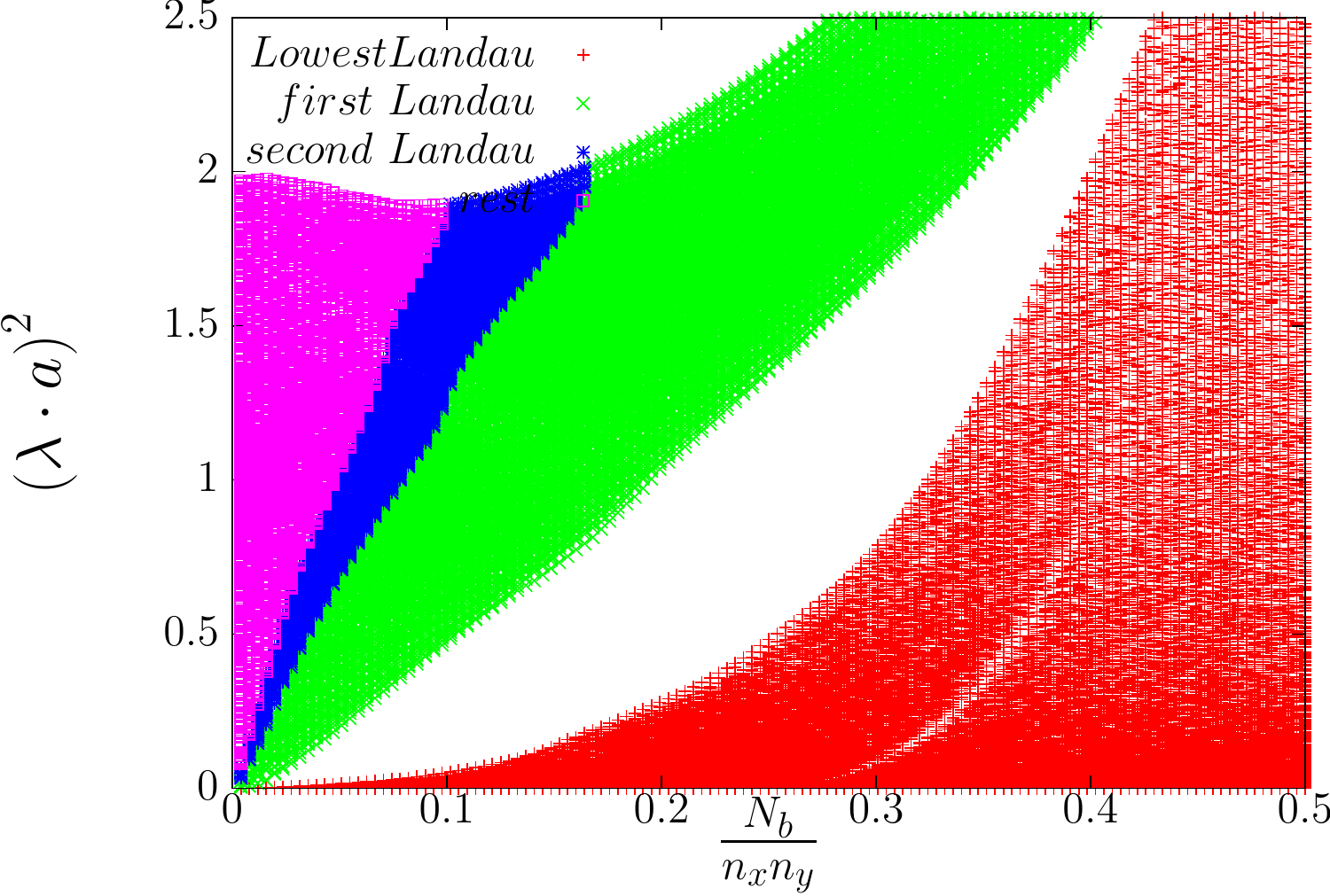}
 \caption{Classification of the lattice eigenvalues
   according to continuum 
Landau-level degeneracies. The left panel shows the spectra of the
free two-dimensional Dirac operator, 
while the right panel shows the interacting case -- evaluated
on a two-dimensional slice of a typical four-dimensional 
gauge configuration.}
\label{fig:gap}
\end{figure}

In order to check this explanation we put the theory on a symmetric
$2d$ lattice $n_{x}=n_{y}$ and  
use the staggered discretization of the Dirac operator. 
Here and in the rest of the paper we only
consider the magnetic field dependence of the operator, i.e., the
valence effect, while setting $B=0$ in the fermion determinant. 
We use $n_{f}=2+1$ flavors of staggered quarks with
physical quark masses. 
On the lattice the degeneracy of the LL-s is broken by lattice
artefacts, and the spectrum forms a fractal structure (Hofstadter butterfly) in
the $\lambda - B$ plane~\cite{Hofstadter:1976zz}.  
This is shown in the left panel of Fig.~\ref{fig:gap}. The different
LL-s are represented by different colors and are identified with the
help of eq.~(\ref{degen}) just by counting modes (and taking into 
account the twofold doubling of eigenvalues in two dimensions).  
%

Next we switch on QCD interactions by taking one 2$d$ $x-y$ slice of a 
typical 4$d$ QCD gauge configuration and inserting the links in the
two-dimensional staggered Dirac-operator  
$\slashed{D}_{xy}$. This smears out 
almost entirely the butterfly structure~\cite{Endrodi:2014vza}
(see the right panel of Fig.~\ref{fig:gap}). 
However, a clear gap is still present, with the number of eigenmodes 
below the gap exactly matching the expected continuum degeneracy 
(\ref{degen}).
These
features enable us to unambigously  
identify the LLL also in two dimensional QCD.
The identification of the 
higher LL-s is however no longer possible after color interactions are
included. 

The reason why 
the LLL survives the presence of strong
interactions is topological. In fact, in 2$d$ and in the continuum,
the index theorem assures that the number of zero modes of $\slashed{D}_{xy}$
equals the magnetic flux, which is a quantized topological invariant,
irrespectively of the presence of SU$(3)$ interactions. 
On the lattice these become almost zero modes, 
which are however protected by topology and remain separated from the
modes above the gap. 
To show that this gap is not just a lattice artefact we have to
perform the continuum limit. 
We take lattices at five different spacings, corresponding to 4$d$
lattices with temporal extent ranging from
$n_{t}=4$ up to $n_{t}=12$, keeping the physical lattice volume and
temperature fixed. To
make a comparison between different lattice spacings,  
we rescale each eigenvalue with the bare quark mass~\cite{Kovacs:2012zq} 
and use the same
physical magnetic field, which  
corresponds to using the same flux $\left(N_b\right)$ on all $n_{t}$-s. 
\begin{figure}[t]
 \centering
 \includegraphics[width=7cm]{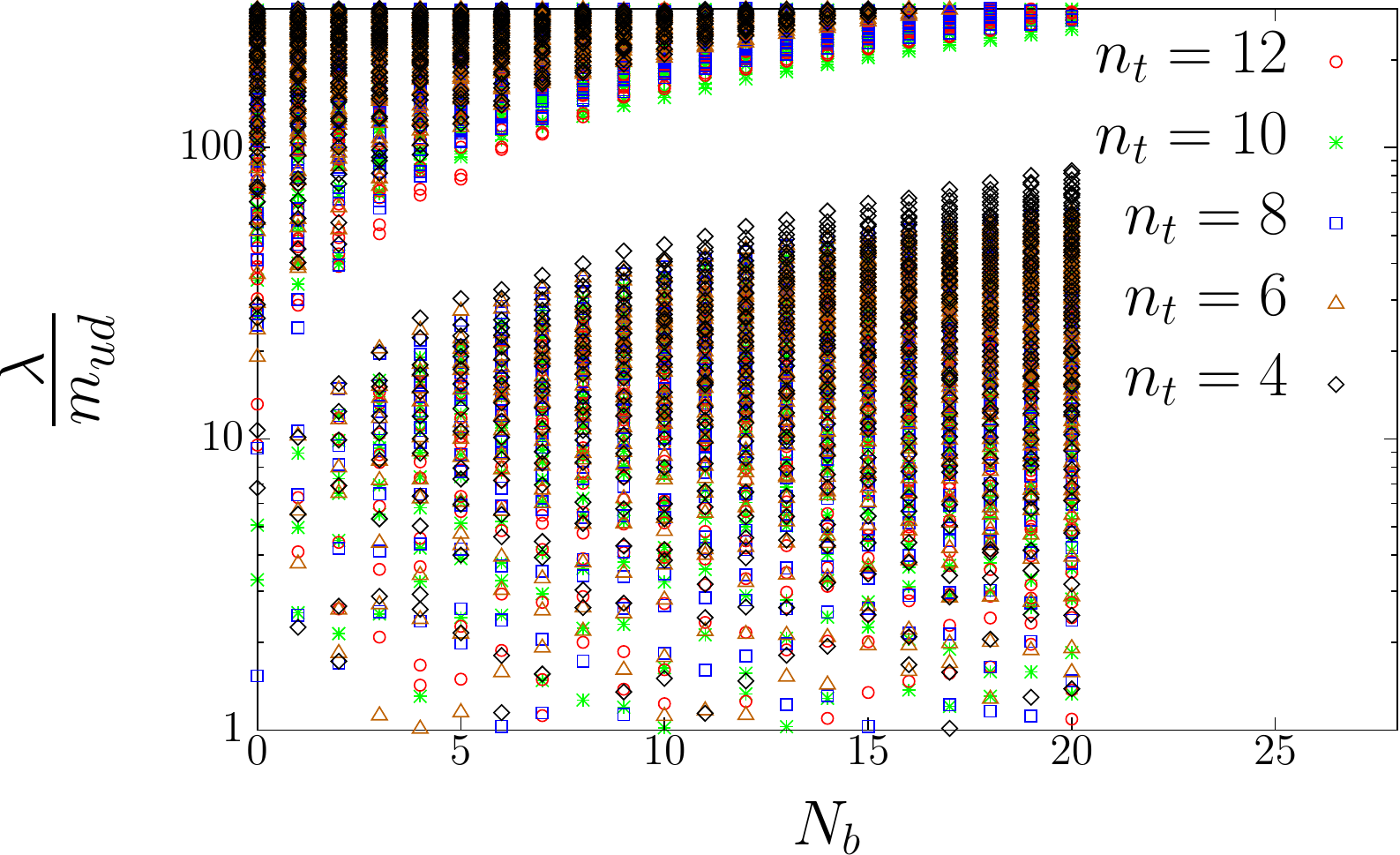}
 \caption{
   The 2$d$ spectrum, rescaled by the  
   bare quark mass, as a function of the magnetic flux, for different
   lattice spacings, ranging from 
   $a=0.125\mathrm{fm}$ to $a=0.041\mathrm{fm}$. 
}
\label{fig:gaplimit}
\end{figure}
We show our results in Fig.~\ref{fig:gaplimit}. The gap between
the first $3N_{b}$ modes and the rest of the spectrum indeed persists
for small flux quanta even at the smallest lattice spacing. 

\section{Lowest Landau level dominance in 2$d$}

\begin{figure}[t]
 \centering
 \includegraphics[width=7cm]{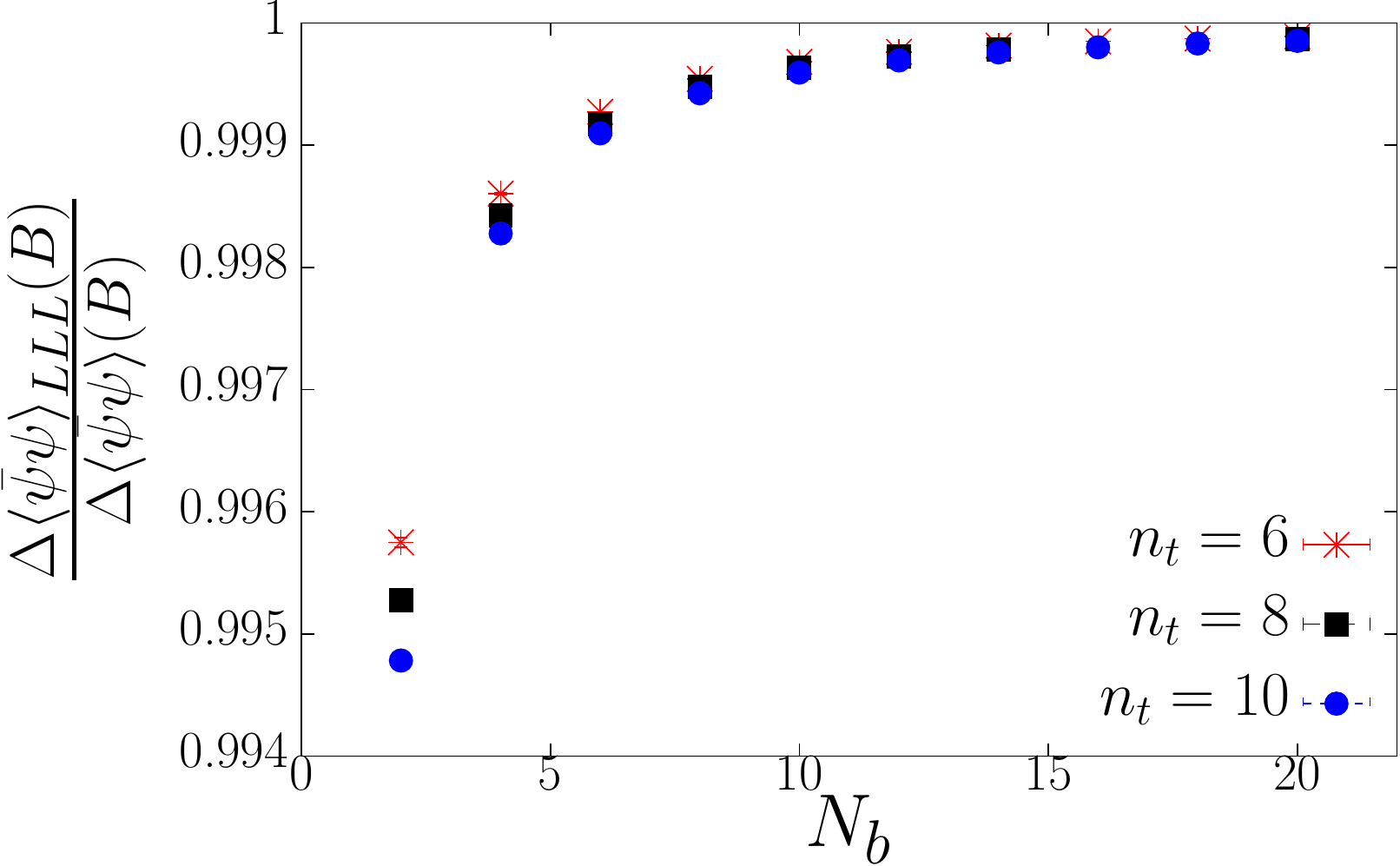}\quad\quad
 \includegraphics[width=7cm]{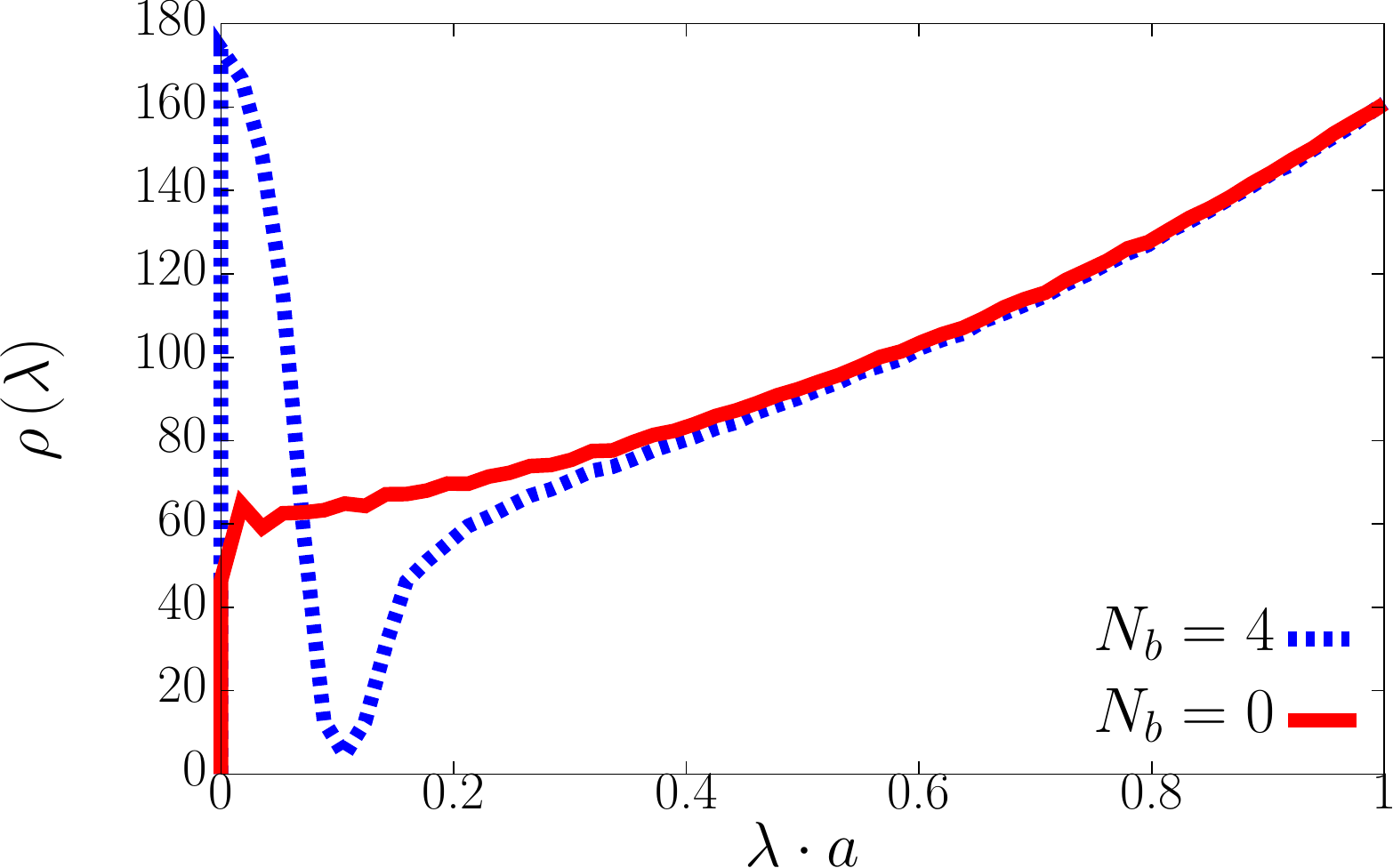}
 \caption{
Left panel: the portion of the change in the condensate 
coming from the LLL at a fixed temperature 
($T$=214 MeV) as a function of the magnetic flux. Right panel: the 
spectral density of the 2$d$ Dirac operator at zero and non-zero external
 magnetic field respectively.
}
\label{fig:dominance}
\end{figure}

Having found a way to identify the LLL in $2d$ QCD, we now turn our
attention to its contribution  
to the valence effect on the chiral condensate discussed in the
Introduction. We begin again with the 2$d$ case, i.e., from the study
of 2$d$ slices of 4$d$ QCD configurations. The contribution from the  
LLL to the condensate is:
\begin{equation}
\label{eq:finiB_2d}
\langle\bar{\psi}\psi\rangle_{B,2d,LLL}=\left\langle\sum_{i\in 1\cdots
  3N_{b}}\frac{2m}{\lambda_{i}(B)^{2}+m^{2}}\right\rangle\,, 
\end{equation}
where $\lambda_{i}(B)$ are the eigenvalues of the finite-$B$ operator.
At zero magnetic field, the contribution of the first $3N_{b}$ modes
to the condensate is
\begin{equation}
\label{eq:zeroB_2d}
\langle\bar{\psi}\psi\rangle_{B=0,2d,LLL}=\left\langle\sum_{i\in 1\ldots
  3N_{b}}\frac{2m}{\lambda_{i}(0)^{2}+m^{2}}\right\rangle\,, 
\end{equation}
where $\lambda_{i}(0)$ are the eigenvalues of the zero-$B$ operator.
The change in the condensate coming from the first $3N_b$ modes, which
at finite $B$ are precisely the LLL modes, is just the
difference between (\ref{eq:finiB_2d}) and (\ref{eq:zeroB_2d}).
To quantify how much of the total change in the condensate comes from
the LLL, we compute the ratio of the differences:
\begin{equation}
\label{2dleveldominance}
\frac{\langle\bar{\psi}\psi\rangle_{B,2d,LLL}-\langle\bar{\psi}\psi\rangle_{B=0,2d,LLL}}
{\langle\bar{\psi}\psi\rangle_{B,2d}-\langle\bar{\psi}\psi\rangle_{B=0,2d}}\,, 
\end{equation}
where $\langle\bar{\psi}\psi\rangle_{B,2d}$ and
$\langle\bar{\psi}\psi\rangle_{B=0,2d}$ are the full condensate at
nonzero and zero $B$, respectively. In the left panel of
Fig.~\ref{fig:dominance} we show 
this ratio as a function of the magnetic flux for three values of the
lattice spacing.  
Remarkably, the LLL explains almost 
entirely the change in the condensate. To further 
illustrate the LLL dominance we show the spectral density of the $2d$
Dirac operator in the right panel of Fig.~\ref{fig:dominance} for zero
and non-zero magnetic field. At non-zero $N_b$ one can clearly see a
drop in the spectral density, corresponding to the gap discussed
above. Moreover, above the drop 
the two spectral densities 
are almost identical, which is the reason why the LLL dominates  
the change in the condensate. 

\section{Landau levels in 4$d$}
After having seen how well the LLL dominance works in $2d$, we generalize
the notion of LL-s to the physically more interesting $(3+1)d$ case. 
Let us first discuss the eigenmodes of the free Dirac operator in a
finite 4$d$ box.  
They factorize into an ($x,y$) dependent part, which is one of 
the 2$d$ LL solutions, and into plane waves describing free propagation 
in the $z$ and $t$ directions. 
The eigenvalues of $-\slashed{D}^2$ are now 
\begin{equation}
\lambda^2= \vert q B\vert \left[2n+1-2s_{z}\mathrm{sgn}\left(qB\right)\right] + 
p_{z}^2 +p_{t}^2\,,~~ n=0,1,\ldots\,,~~ s_{z}=\pm\frac{1}{2},
\end{equation}
where
$p_z=\frac{2\pi}{L_{z}}k_z$ and $p_t=\frac{\pi}{L_{t}}\left(2k_t+1\right)$,
with integer $k_z$ and $k_t$, are the momentum in the $z$ and $t$
directions, respectively. The gap, 
which was present in the 2$d$ case, is now filled due to the
contribution of the $z$ and $t$ momenta.
Thus by just looking at the spectrum
we are no longer able to identify the LLL. The situation is obviously
even more complicated after SU$(3)$ interactions are switched
on. However, we have seen that also in this case we can identify the
LLL among the $2d$ modes on an arbitrary ($z,t$) slice. Now naturally arises
the question whether these LLL $2d$ modes have any special  
role in the $4d$ spectrum.
To answer this question, we begin by determining what is the
overlap of a $2d$ eigenmode 
with a given $4d$ mode, i.e.,
\begin{equation}
\label{wweight}
W_{j}\left(B\right)=\sum_{z,t}\vert \left(\phi_{t,z,j}\left(B\right),
  \psi \right) \vert^2, 
\end{equation} 
where the summation is over all ($z,t$) slices,
$\phi_{t,z,j}\left(B\right)$ is the $j$-th positive eigenmode of the $2d$ Dirac
operator on the ($z,t$) slice and $\psi$ is a $4d$ eigenmode.
\begin{figure}[t]
 \centering
 \includegraphics[width=7cm]{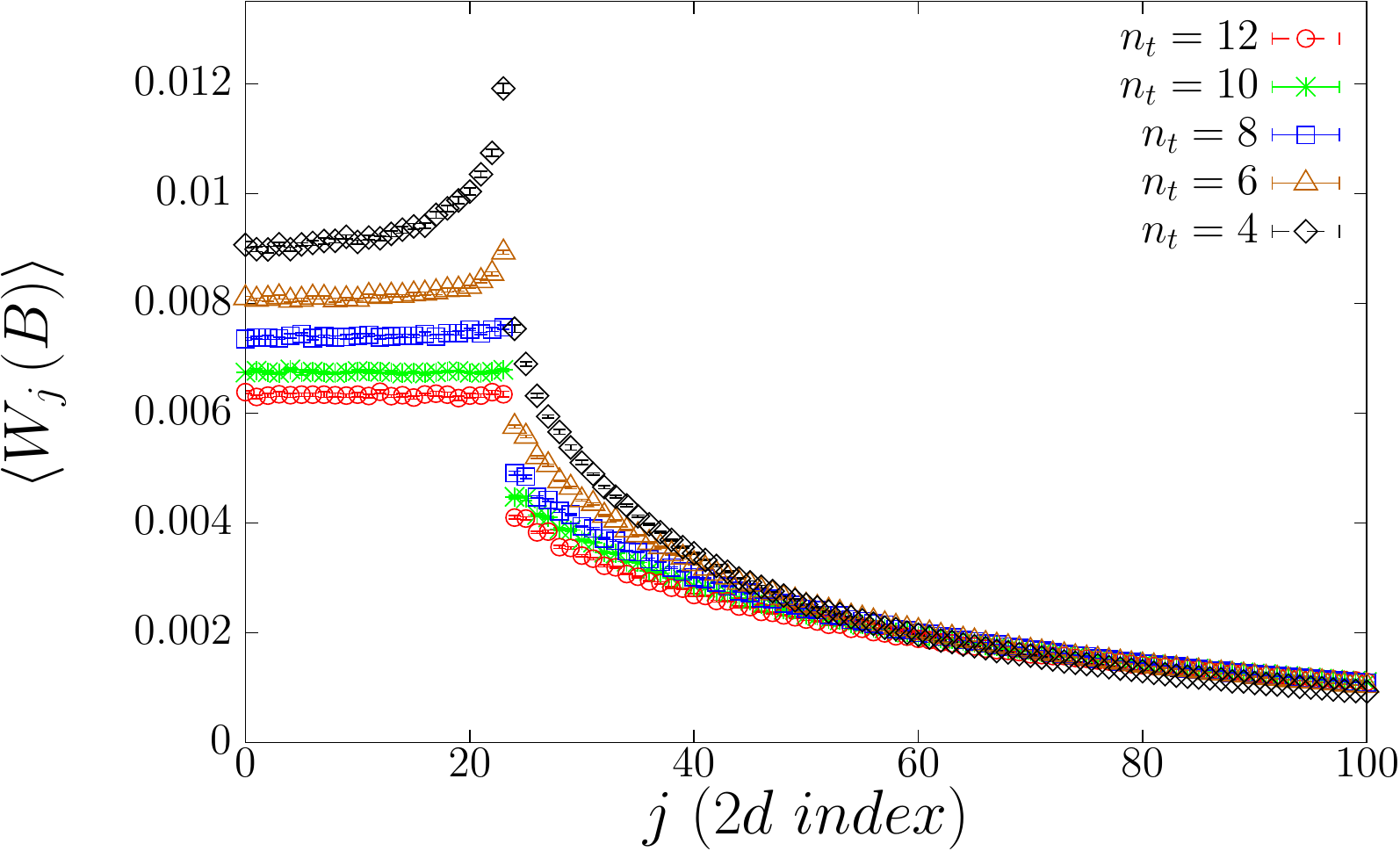}\hfil
 \includegraphics[width=7cm]{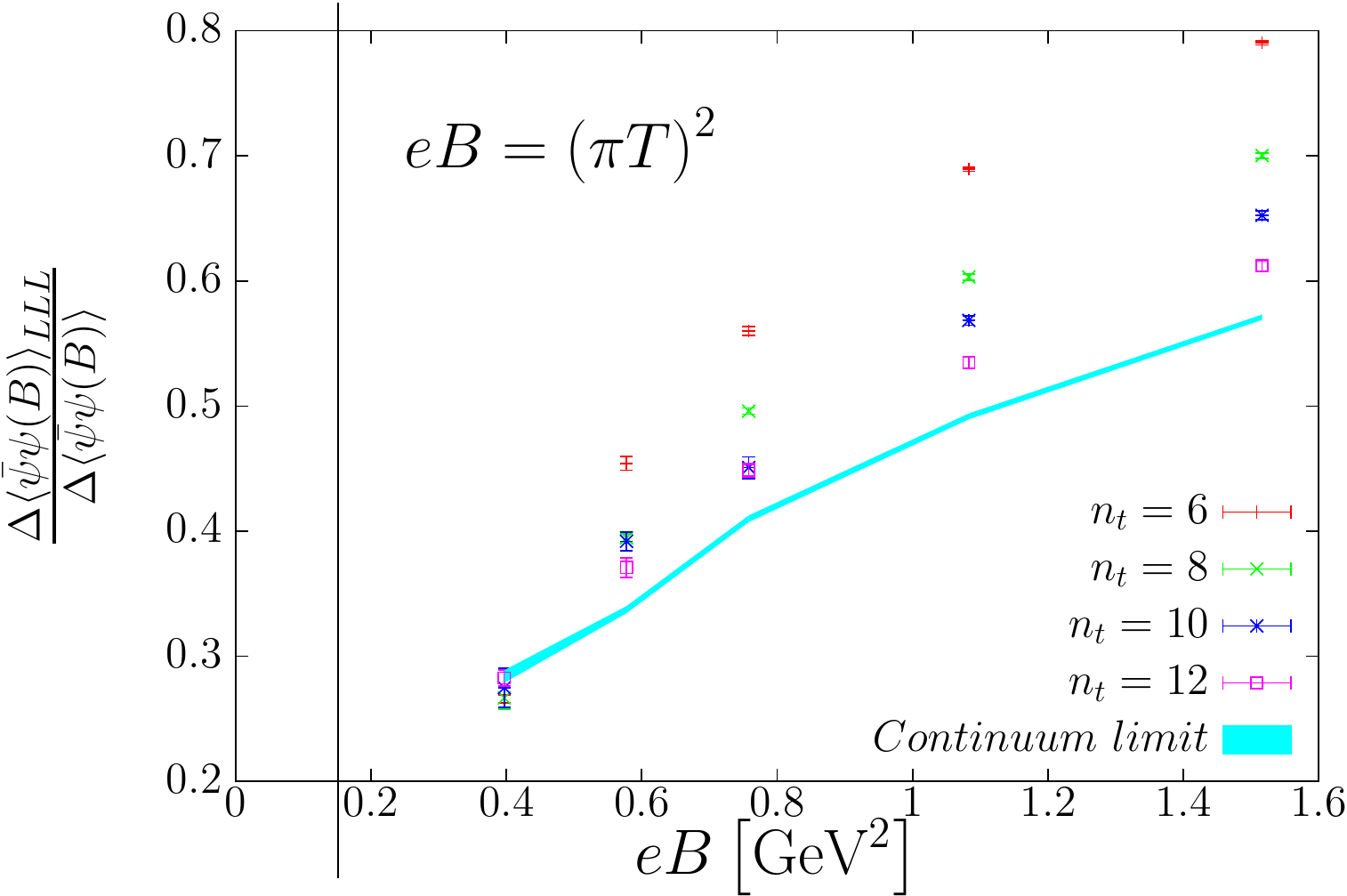}
 \caption{Left panel: overlap between a 4$d$ mode and the $j$-th $2d$
   eigenmode, as defined in (\protect\ref{wweight}). An average over $4d$
   eigenmodes in the range 
   $\frac{\lambda}{m_{ud}}\in\left[220,225\right]$, and an average over
   several gauge configurations are performed. Right panel:
The fraction of the change in the valence condensate in 4$d$ that can 
be attributed to the change of the first $3N_b$ modes on all ($z,t$)
slices as a function 
of the physical magnetic field. The temperature is $124$ MeV. 
 } 
\label{fig:weightlimit}
\end{figure} 

Our results are shown in the left panel of Fig.~\ref{fig:weightlimit} for
$N_{b}=8$ and $T\simeq 400\mathrm{MeV}$. Here we averaged over modes in a 
small spectral window in the low part of the $4d$ spectrum and over gauge configurations.
We can see that there is a jump in $W_j$ after the first $3N_b$ modes, which
make up the LLL in the $2d$ case. Thus the LLL gives indeed a distinct
contribution to the $4d$ modes, and it seems therefore reasonable to
define the LLL contribution to the 4$d$ eigenmodes by projecting them
onto the 2$d$ LLL modes.


\section{Lowest Landau level dominance in 4$d$}
Having defined the LLL part of a $4d$ eigenmode we can determine the
LLL contribution to  
the valence effect. The change in the condensate in $4d$ due to the
appearance of $3N_{b}$  
would-be zero modes on each $z,t$ slice can be calculated in a similar
manner as in the $2d$  
case. The contribution to the condensate from the LLL modes at finite
$B$ is identified as
\begin{equation}
\langle\bar{\psi}\psi\left(B\right)\rangle_{3N_{b},~all~slices} 
= \left\langle\sum_{i}\frac{2m}{\lambda_{i}(B)^{2}+m^{2}} C_i(B)\right\rangle\,,
\end{equation}
where $C_i(B)$ is the size of the projection of mode $i$ on the LLL
subspace, $C_{i}\left(B\right)=\sum_{doublers}\sum_{j=1}^{3N_b}W_j$,
where the summation includes the negative doublers of the 2$d$ LLL
modes.\footnote{The $i$ dependence of $C$ comes from the fact that the 
  scalar product in $W_j$ (\ref{wweight}) has to be evaluated with
  $\psi_i$.} In order to calculate the change in the condensate due
to these modes we have to subtract the contribution of the first
$3N_{B}$ $2d$ modes at $B=0$: 
\begin{equation}
\langle\bar{\psi}\psi\rangle_{3N_{b},~all~slices} 
= \left\langle\sum_{i}\frac{2m}{\lambda_{i}(0)^{2}+m^{2}} C_i^0(N_b)\right\rangle\,.
\end{equation}
Here $C_i^0(N_b)$ is the size of the projection of mode $i$ on the
subspace built out of the first $3N_b$ 2$d$ modes on each slice,
\begin{equation}
  \label{eq:proj0}
  C_i^0(N_b) = \sum_{doublers}\sum_{t,z}\sum_{j=1}^{3N_b} |(\phi_{t,z,j}(0),\psi_i)|^2\,,
\end{equation}
where $\phi_{t,z,j}(0)$ is the $j$-th 2$d$ mode on slice $(t,z)$ in the absence 
of the magnetic field.
In the right panel of Fig.~\ref{fig:weightlimit} 
we show what fraction of the valence effect comes from
the fact that the nature of the first $3N_b$ modes on all ($z,t$)
slices has changed. 
The continuum limit has been obtained by assuming that 
$\frac{\Delta\langle\bar{\psi}\psi\left(B\right)\rangle_{LLL}}{\Delta\langle\bar{\psi}     
\psi\left(B\right)\rangle}$
approaches it quadratically in the lattice
spacing, which leads to an acceptable
$\chi^2$ for all values of the magnetic field. The contribution of the
2$d$ LLL modes gets larger and larger as the magnetic field increases,
which suggests LLL dominance for large $B$, as expected from effective models.
In fact they use the LLL approximation in the limit $qB \gg  
\left(\pi T\right)^2$. We show the magnetic field corresponding to
$\left(\pi T\right)^2$ 
with a vertical line. We conclude that for our largest magnetic
field 50$\%$ of the 
valence effect comes from the LLL in the physical 4d case. In the
future it will be very interesting to see whether the ``sea'' effect
can also be reproduced in the LLL approximation. 

\end{document}